%% file: daleknn1.tex
\documentclass[usenatbib, twocolumn]{aastex63}

\usepackage{newtxtext,newtxmath}

\usepackage[T1]{fontenc}
\usepackage{ae,aecompl}
\usepackage{comment}
\usepackage{booktabs}

\usepackage{savesym}
\savesymbol{tablenum}
\usepackage{siunitx}
\restoresymbol{SIX}{tablenum}

\usepackage{graphicx}	
\usepackage{amsmath}	
\usepackage{amssymb}	

\input{glossary}

\usepackage[textsize=footnotesize,color=white,bordercolor=red,linecolor=red]{todonotes}

\newcommand{\zenododata}{\url{https://tinyurl.com/y77gokrr}}
\newcommand{\tardis}{\gls{tardis}\xspace}

\submitjournal{ApJ}
\shorttitle{\textsc{Dalek}}
\shortauthors{Kerzendorf et al.}

\begin{document}
\title{\textsc{Dalek} -- a deep-learning based emulator for \textsc{tardis}}
\correspondingauthor{Wolfgang E. Kerzendorf}
\email{wkerzendorf@gmail.com, wkerzend@msu.edu}
\author[0000-0002-0479-7235]{Wolfgang E. Kerzendorf}
\affiliation{Department of Physics and Astronomy, Michigan State University, East Lansing, MI 48824, USA}
\affiliation{Department of Computational Mathematics, Science, and Engineering, Michigan State University, East Lansing, MI 48824, USA}

\author[0000-0002-7941-5692]{Christian Vogl}
\affiliation{Max-Planck-Institut f\"ur Astrophysik, Karl-Schwarzschild-Str. 1, D-85741 Garching, Germany}

\author[0000-0003-0426-6634]{Johannes Buchner}
\affiliation{Max-Planck-Institut f\"{u}r extraterrestrische Physik, Giessenbachstrasse 1, 85748 Garching bei M\"{u}nchen, Germany}

\author[0000-0002-3011-4784]{Gabriella Contardo}
\affiliation{Center for Computational Astrophysics, Flatiron Institute, New York, NY, 10010, USA}

\author[0000-0003-2544-4516]{Marc Williamson}
\affiliation{Department of Physics, New York University, New York, NY, 10003, USA}

\author[0000-0003-4418-4916]{Patrick van der Smagt}
\affiliation{Machine Learning Research Lab, Volkswagen AG, Munich, Germany}
\affiliation{Faculty of Informatics, E\"otv\"os Lor\'and University, Budapest, Hungary}

\begin{abstract}
Supernova spectral time series contain a wealth of information about the progenitor and explosion process of these energetic events. The modeling of these data requires the exploration of very high dimensional posterior probabilities with expensive radiative transfer codes. Even modest parametrizations of supernovae contain more than ten parameters and a detailed exploration demands at least several million function evaluations. Physically realistic models require at least tens of CPU minutes per evaluation putting a detailed reconstruction of the explosion out of reach of traditional methodology. 
The advent of widely available libraries for the training of neural networks combined with their ability to approximate almost arbitrary functions with high precision allows for a new approach to this problem. Instead of evaluating the radiative transfer model itself, one can build a neural network proxy trained on the simulations but evaluating orders of magnitude faster. Such a framework is called an emulator or surrogate model. 
In this work, we present an emulator for the \textsc{tardis} supernova radiative transfer code applied to Type Ia supernova spectra. We show that we can train an emulator for this problem given a modest training set of a hundred thousand spectra (easily calculable on modern supercomputers). The results show an accuracy on the percent level (that are dominated by the Monte Carlo nature of \textsc{tardis} and not the emulator) with a speedup of several orders of magnitude. This method has a much broader set of applications and is not limited to the presented problem.

\end{abstract}




\section{Introduction}

Supernova spectra arise from a complex interplay of processes. Simulating them self-consistently is a computationally intensive endeavour ranging from single simulations taking several CPU minutes to  thousands of CPU hours on large supercomputers. 

The \tardis supernova spectrum synthesis code can  evaluate a single parametrized explosion model within $\approx10$~CPU~minutes with some approximations that have a minor impact on the output. One of the goals of \tardis is to perform a Bayesian parameter inference on spectral time series. However, even for a very simple model for a single supernova spectrum with a fixed density profile and ten uniform abundances this results in a  dozen parameters. Such parameter spaces require millions of evaluations for parameter searches which is infeasible even for fast codes like \tardis. 

Emulators are a solution to this problem \citep[see][for an early implementation of emulators]{2015apj...812..128c}. These constructs approximate simulations by using functions that are easy to fit 
to a grid of simulations and are fast to evaluate.
\citet{stefan_lietzau_2017_1312512} did attempt to emulate \tardis using \gls{pca} and \gls{gp} regression. \citet{stefan_lietzau_2017_1312512}'s emulator worked on eleven abundances for \snia simulations with \tardis but was not able to work on the full set of thirteen parameters. \citet{2020A&A...633A..88V} showed that using \gls{pca} and \gls{gp} emulator technique worked for the lower (five) dimensional  space of \gls{sniip} spectra. 

Neural Networks have been shown to be universal function approximators \citep{cybenko1989,HORNIK1989359}. 
\tardis can be seen as a function that takes an input vector of parameters and transforms these into a spectral vector. We emulate an equivalent parameter space (our parameter space taking nuclear decay into account) to the work of \citep{stefan_lietzau_2017_1312512} using neural networks.

In Section~\ref{sec:methods}, we describe the methods used in this emulation attempt. Section~\ref{sec:results} summarizes the performance of the emulator (accuracy and computational efficiency).
We conclude the paper in Section~\ref{sec:conclusion} and give an outlook of future work.

\section{Methods}
\label{sec:methods}

The aim of the proposed emulator is to explore the posterior of Type Ia supernova spectra at roughly ten days before maximum using a uniform model for the abundances. We varied the abundances of 9 elements, one isotope, the velocity of the inner boundary, and temperature of the inner boundary \citep[see][ for 
a description of these parameters]{2014MNRAS.440..387K}. All other parameters of the model remain fixed. We chose the density profile \texttt{branch85\_w7} \citep[a powerlaw density profile; see ][for details]{2014MNRAS.440..387K} and an outer boundary velocity of \num{20000}\,\kms. The plasma calculation used the \texttt{nebular} setting for ionization and \texttt{dilute-lte} setting for excitation. We use the formal integral calculated spectrum for our emulation purposes (see \url{https://tardis-sn.github.io/tardis/physics/montecarlo/sourceintegration.html}). The input \textsc{tardis} configuration file is available in the data cache linked to this paper \zenododata.

There are several steps to construct an emulator for \tardis: 1) Selecting the training set of parameters covering the necessary parameters for the specific problem. 2) Calculating the \tardis spectra for the training set. 3) Constructing a neural network architecture. 4) Training the neural network architecture.

\subsection{Data set}

We are trying to use a parameter space that is close to physically realistic values. \sn{2002}{bo} is one of the most well studied \sneia (309 results in ADS). \citet{2005mnras.360.1231s} have done a detailed abundance tomography on this object, and \citet{2011phdt.......324k} used this object for initial automated fitting attempts, which 
are a pre-cursor to the work presented here. The assumptions made in \tardis make it most accurate before maximum and we will focus on 8.9 days after explosion (roughly ten days before maximum). We divide the creation of the training set into finding suitable abundance combinations and finding suitable inner boundary velocity and temperature combinations. 

\citet{2005mnras.360.1231s} model the spectrum at $8.9\,d$ after exposion using $v_\textrm{inner}=\num{13900}\,\kms$ and $T_\textrm{inner}=\num{11850}\,\textrm{K}$. We construct a uniformly spaced training grid with inner boundary temperatures ($T_{inner}$) between 10000-14000 K and inner velocities ($v_{inner}$) between 10000-15000 km/s. This grid safely contains the accepted values of the parameters presented in \citet{2005mnras.360.1231s}.

We rely on theoretical nucleosynthesis calculations
given in the Heidelberg Supernova Model Archive (HESMA; \url{https://hesma.h-its.org})
to find physically viable abundances.
We use 62 spherically averaged isotopic models \citep[presented in the following papers;][]{2010Natur.463...61P,2010A&A...514A..53F,2010ApJ...714L..52S,2010ApJ...719.1067K,2012ApJ...747L..10P,2012MNRAS.420.3003S,2012ApJ...750L..19R,2013MNRAS.429.1156S,2013MNRAS.429.2287K,2013A&A...554A..67S,2013MNRAS.436..333S,2013ApJ...778L..18K,2014MNRAS.438.1762F,2014A&A...572A..57O,2015MNRAS.450.3045K,2015A&A...580A.118M,2016A&A...592A..57S,2016MNRAS.459.4428K,2017MNRAS.472.2787N,2018A&A...618A.124F} for the creation of the training set (the online data \zenododata\ contains the specific list of models). We only use abundances that are in cells with velocities above \num{10000}\,km/s to be self-consistent with our choice of inner boundary velocities. 

 The training set is created with the abundances of O, C, Mg, Si, S, Ca, Ti, Cr, Fe (stable), and \nucl{Ni}{56}. We then calculate the location of the 20\% and 80\% quantile for each element excluding oxygen. We sample uniformly in $\log_{10}$-space between these two quantiles for all elements. Finally, we set the oxygen abundance to the remaining part of the abundance fraction. .
\begin{table*}
    \input{training_grid_description.tex}
    \label{tab:training_set_overview}
    \caption{The distributions of elemental abundances (uniformly in $\log_{10}$-space) for the training set. The two additional parameters have uniform distributions: $\textrm{T}_\textrm{inner} = 10000 - 14000\,\textrm{K}$ and $\textrm{v}_\textrm{inner}=10000 - 15000\,\kms$.%
    }
    
\end{table*}

We removed any combination of these parameters that would lead to an input luminosity of less than $1\times10^{42}\,\textrm{erg}\,\textrm{s}^{-1}$. 
The extent of the training parameter set can be seen in Table~\ref{tab:training_set_overview}.

We experimented with several choices of number of packets for each Monte Carlo iteration and gauged the variation for the spectrum creation resulting from the Monte Carlo nature of \tardis \citep[see][for details of this process]{2014MNRAS.440..387K}. The choice of \num{100000} Monte Carlo packets for each iteration (opting for 30 iterations in total and increasing the number to \num{200000} packets for the last iteration) resulted in spectra that had less than 1\% intrinsic noise---far lower than the systematic uncertainties present in the comparison between data and spectra. 

We calculated a training/validation data set with \num{98000} samples and a test set with \num{19930} samples on the MSU high-perfomance cluster provided by the Institute for Cyber-Enabled Research.

We resampled the spectra from \tardis on a logarithmic grid between $3400\,\AA$ and $7600\,\AA$ to make the line structures across the spectrum have roughly equal pixels per structure. The final data set has 12 input parameters and 500 spectral data points. The input \tardis file, parameters (abundances, inner boundary velocity/temperature), and spectra are available at \zenododata.

\subsection{Neural network architecture \& training }

We split the group of \num{98000} spectra into a set of \num{68600} (=70\%) for training the neural networks and \num{29400} for cross-validation.
Each data point consisted of 12 inputs (the `parameters') and 500 outputs (the `spectra').
Both input and output values were preprocessed by first taking the $\log_{10}$ of the values, after which the values were normalised by removing the mean and scaling to unit variance with  \textsc{StandardScaler} \citep[\textsc{SciKit-Learn};][]{scikit-learn}.

We use feed-forward neural networks to efficiently approximate and generalize these data.
Even though training a neural network may cost a few hours of computation time, inference with trained neural networks is very fast since it only involves a small number---for the architectures used in this paper in the order of $10^6$---of floating-point operations and a few hundred nonlinear function evaluations.  

We trained a number of feed-forward neural networks of different topology on the data.
The neural networks were implemented in \textsc{Keras} on \textsc{TensorFlow} 1.14 or 2.0.

Good neural network architectures were found by hyperparameter search.
We used cluster-based hyperparameter search using Polyaxon 0.5.6 (\url{https://polyaxon.com/}) on a cluster of IBM and Nvidia machines, each with multiple Tesla V100 GPUs.
Training a single neural network lasts 4--7 hours on such an architecture, and we parallelized over 200 instances.

Table~\ref{tab:hyper_param} lists the hyperparameters over which we searched, and their range of possible values.

\begin{table}
\input{hyper_param_table}
\end{table}

We chose to train the network for \num{15000} epochs for networks trained without dropout and \num{40000} epochs with dropout.
Both numbers were chosen with a considerable margin.
From the approximately \num{4000} runs we selected the best results by analyzing their average loss (using mean squared error) over the cross-validation data set.
The best found neural network architectures had a width of 200 to 400 neurons in one or two hidden layers, a softplus activation function, and Nesterov-adam as optimizer.
Dropout never improved the results; batch normalization was not among the best 10 but in the best 50 networks (7\% worse).
As activation function softplus was in the top 30, but the difference in error with neural networks with elu, relu, selu, or tanh activations functions was not more than about 3\%.
The used batch size had little influence, nor did the choice of initializer.

\begin{table}[ht!]
\centering
    \input{ensemble_list.tex}
    \caption{List of five best neural network architectures used in the ensemble training.}
    \label{tab:ensemble_networks}
\end{table}

We then selected the best neural networks for ensemble modeling \citep{opitz1999popular}.
The selected network architectures, those with the lowest loss on the cross-validation data, are listed in Table~\ref{tab:ensemble_networks}.
Ensemble modeling was done by averaging over all listed neural networks.

\section{Results}
\label{sec:results}

In the following, we used the predictions of the ensemble neural network when comparing with the \tardis spectra from the test set (unless otherwise noted).
We used both the maximum fractional error and mean fractional error metrics \citep[see also][]{2020A&A...633A..88V} for comparisons:

\begin{eqnarray}
    \textrm{MeanFE} &= \frac{1}{N}\sum_{i=0}^N 
                \frac{|f_{\lambda,i}^\textrm{emu}-f_{\lambda,i}^\textrm{test}|}{f_{\lambda, i}^\textrm{test}}\\
    \textrm{MaxFE} &= \max_{i=0}^N 
        \frac{|f_{\lambda,i}^\textrm{emu}-f_{\lambda,i}^\textrm{test}|}{f_{\lambda, i}^\textrm{test}}
\end{eqnarray}
with $N$ being the number of pixels in our spectra (in our case 500), $f_{\lambda, i}$ the flux at the $i$-th pixel in the test set. For the training of the emulator we chose to use spectra in $\log_{10}$. However, for the evaluation of the emulator, we will use the linear space as any likelihood comparing the emulated spectrum to an observed spectrum will be in linear flux units.

\begin{figure*}
\centering
\includegraphics[width=0.48\textwidth]{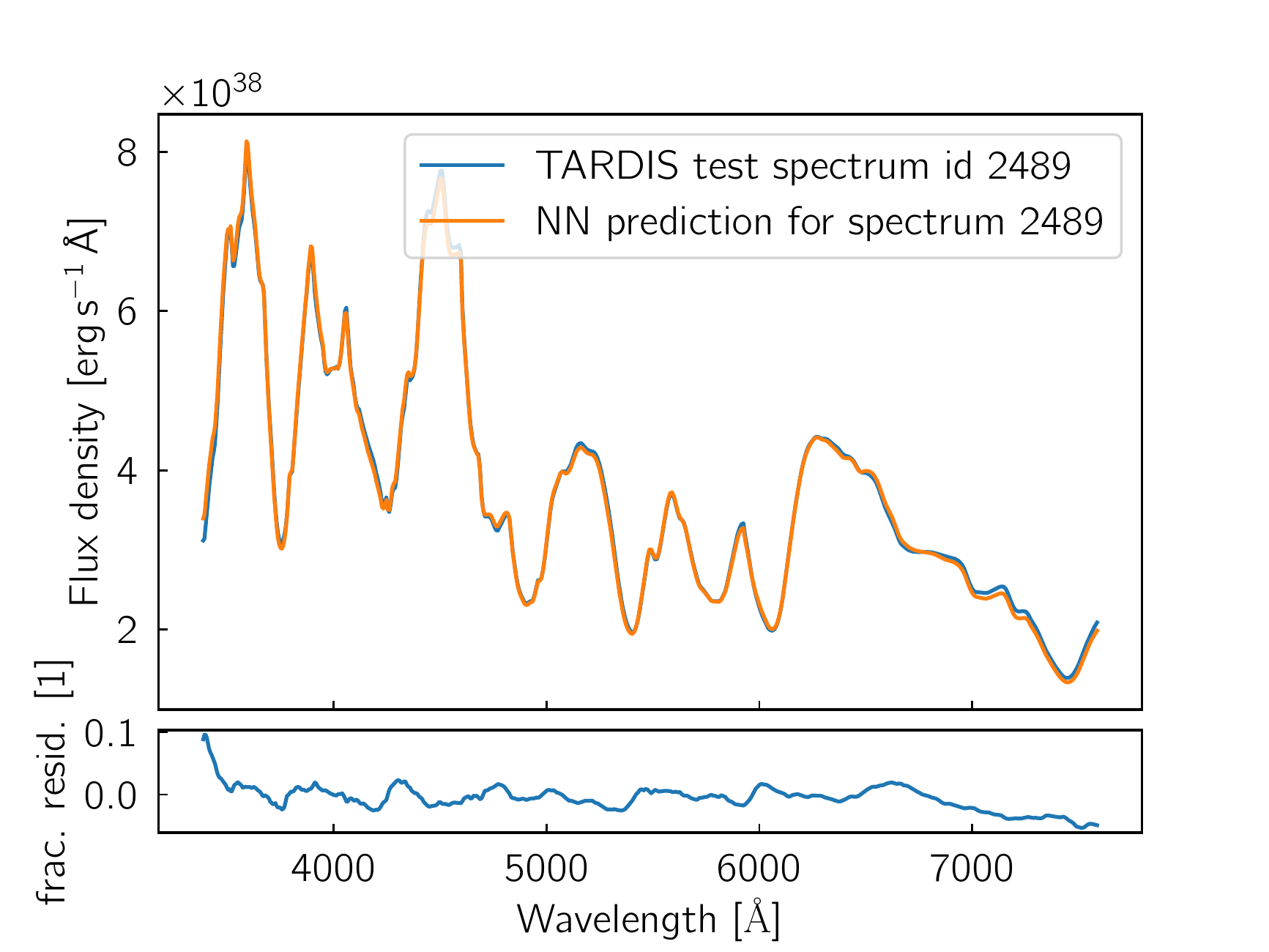}
\includegraphics[width=0.48\textwidth]{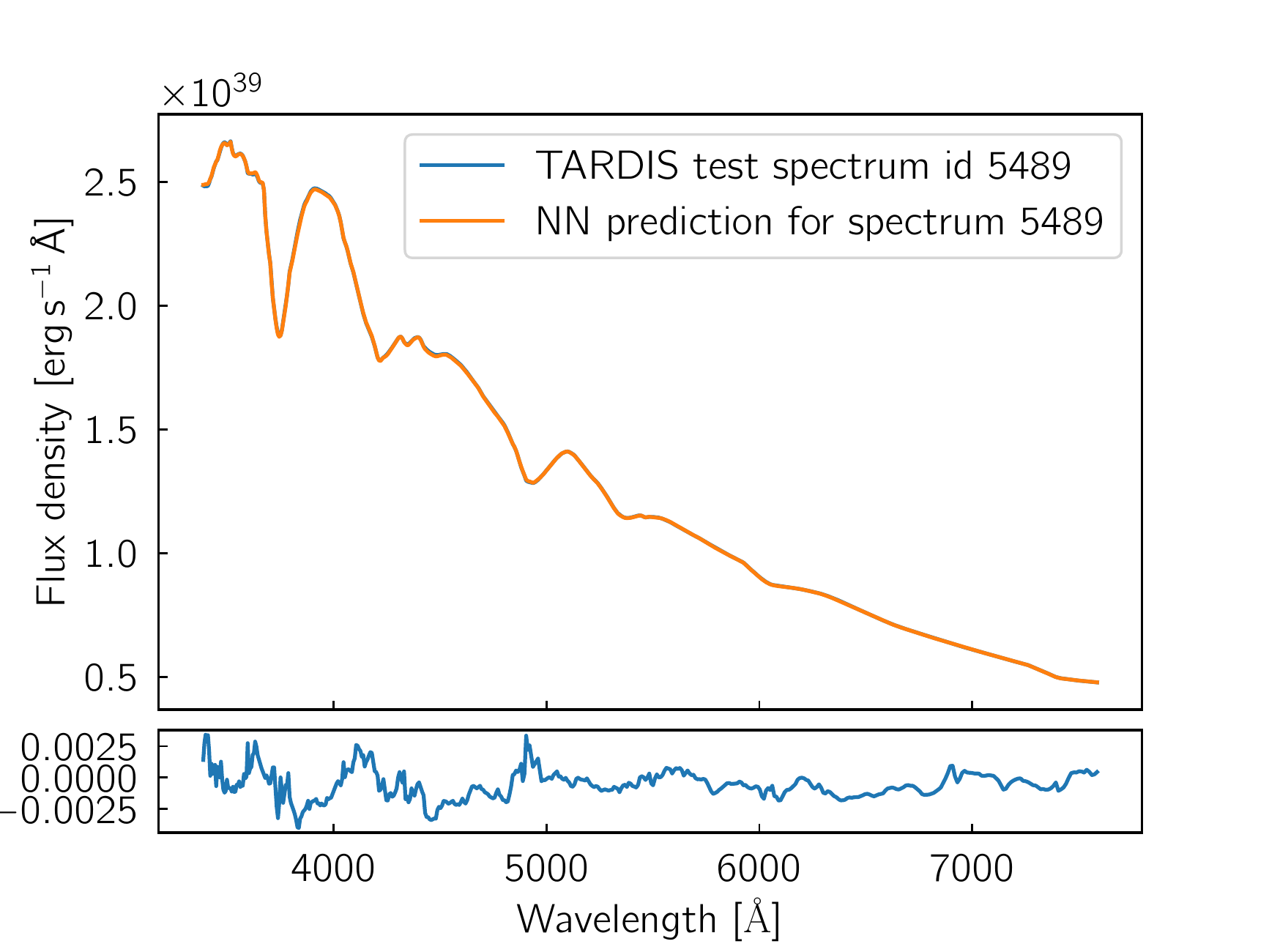}
\caption{Comparison of a spectrum from the test set with a prediction from the ensemble emulator. We showcase the spectra with the highest and the lowest MaxFE from the set of test set predictions. \textbf{Left:} Largest MaxFE from the test set ($\approx 10\%$). \textbf{Right:} Smallest MaxFE from the test set ($\approx 0.4\%$)}
\label{fig:best_worst_spectrum}
\end{figure*}

The ensemble neural network emulator performs well in both metrics with 99\% of predictions having a $\textrm{MaxFE}<0.049$ and $\textrm{MeanFE}<0.014$ and a median prediction of $\textrm{MaxFe}=0.016$ and $\textrm{MeanFE}=0.004$. Figure~\ref{fig:best_worst_spectrum} shows the best and worst prediction in the test set including residuals.

\begin{figure*}
    \centering
    \includegraphics[width=0.48\textwidth]{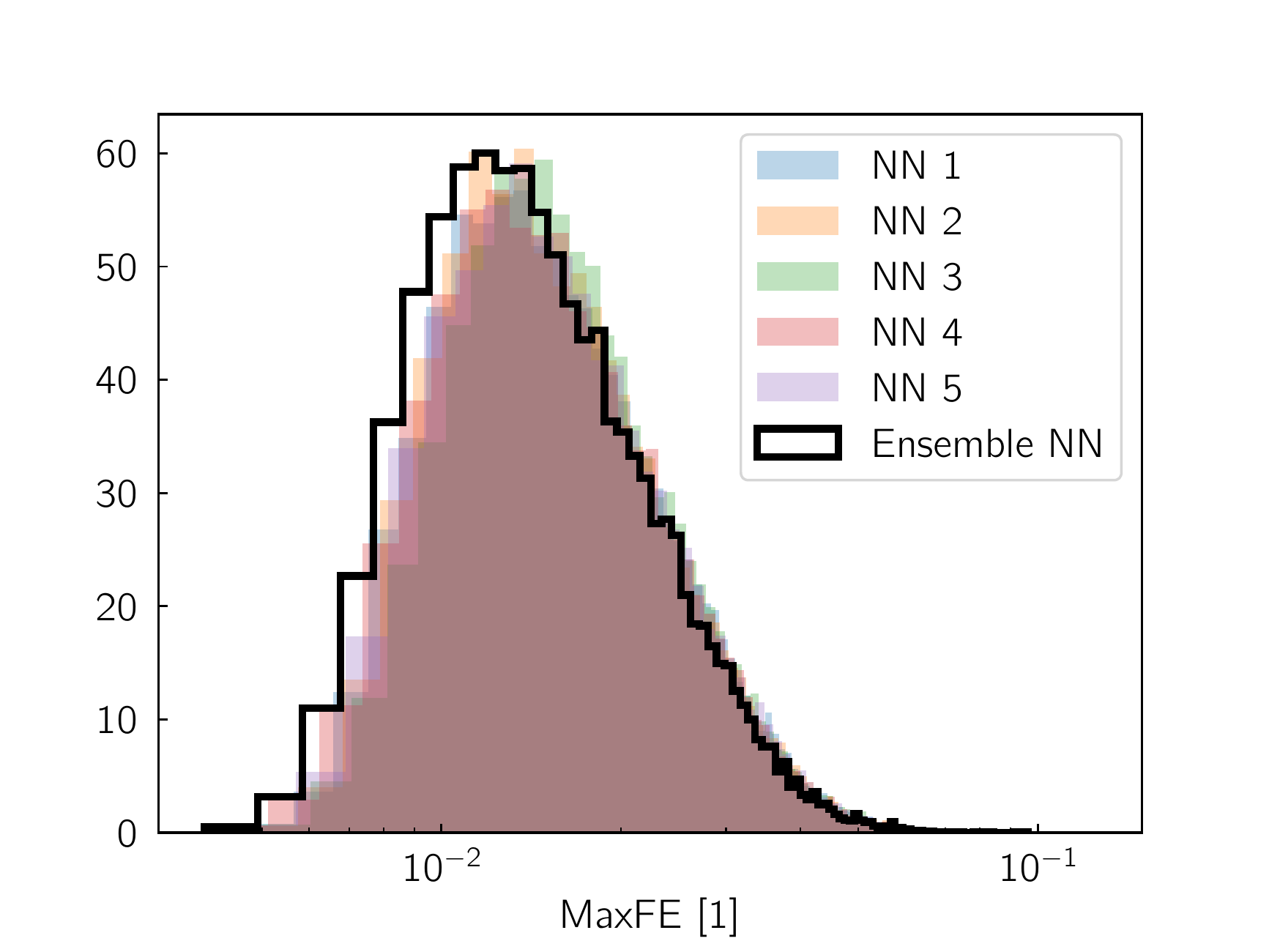}
    \includegraphics[width=0.48\textwidth]{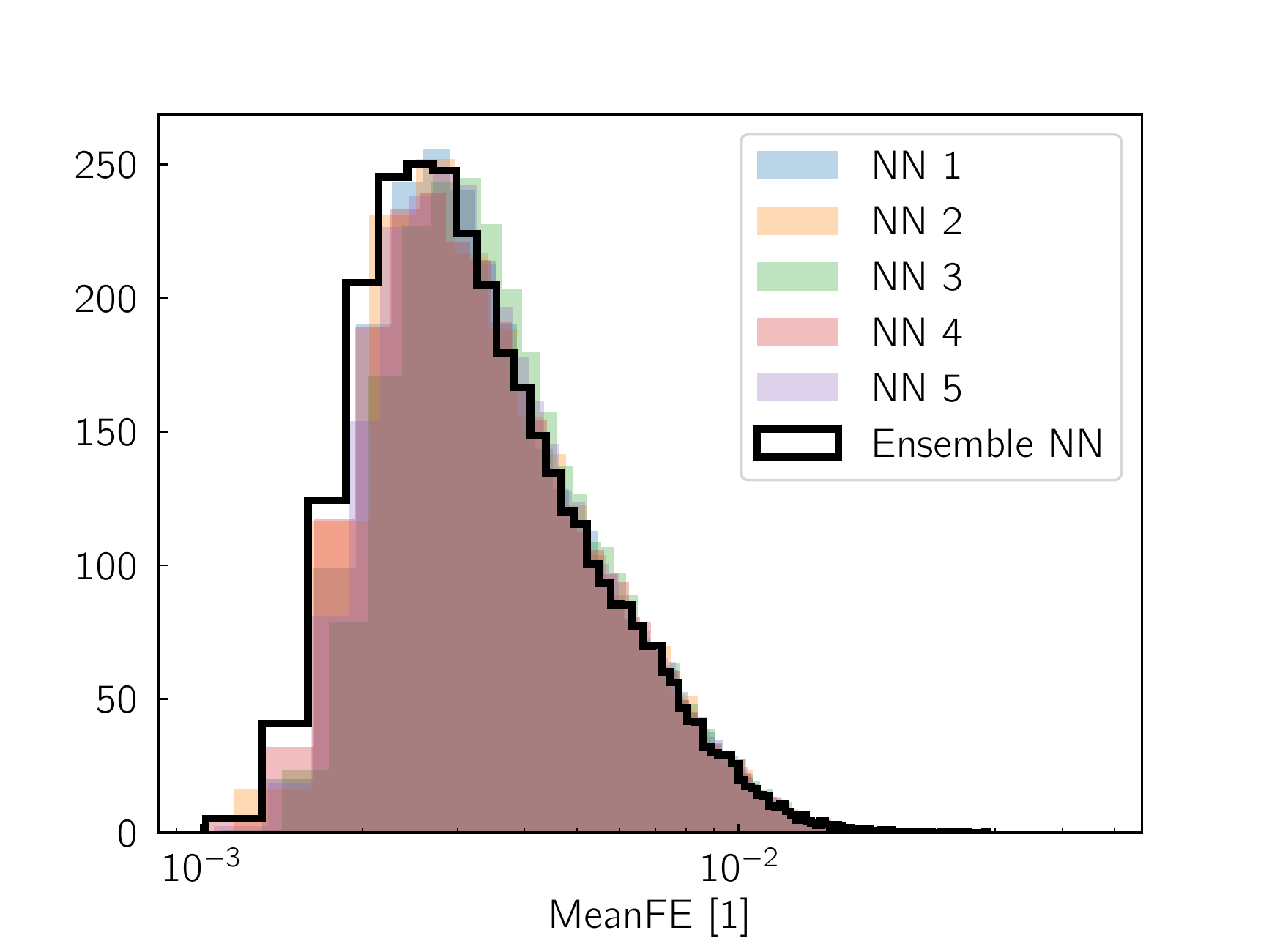}
    \caption{Histogram of prediction uncertainties for the test set using the MaxFE metric on the left and the MeanFE metric on the right.}
    \label{fig:ensemble_uncertainty}
\end{figure*}
Figure~\ref{fig:ensemble_uncertainty} shows the distribution and also compares the prediction uncertainty to the networks that make up the ensemble. The ensemble has roughly a $10\%$ improvement in $\textrm{MeanFE}$ over the individual networks.

\begin{figure}
    \centering
    \includegraphics[width=\columnwidth]{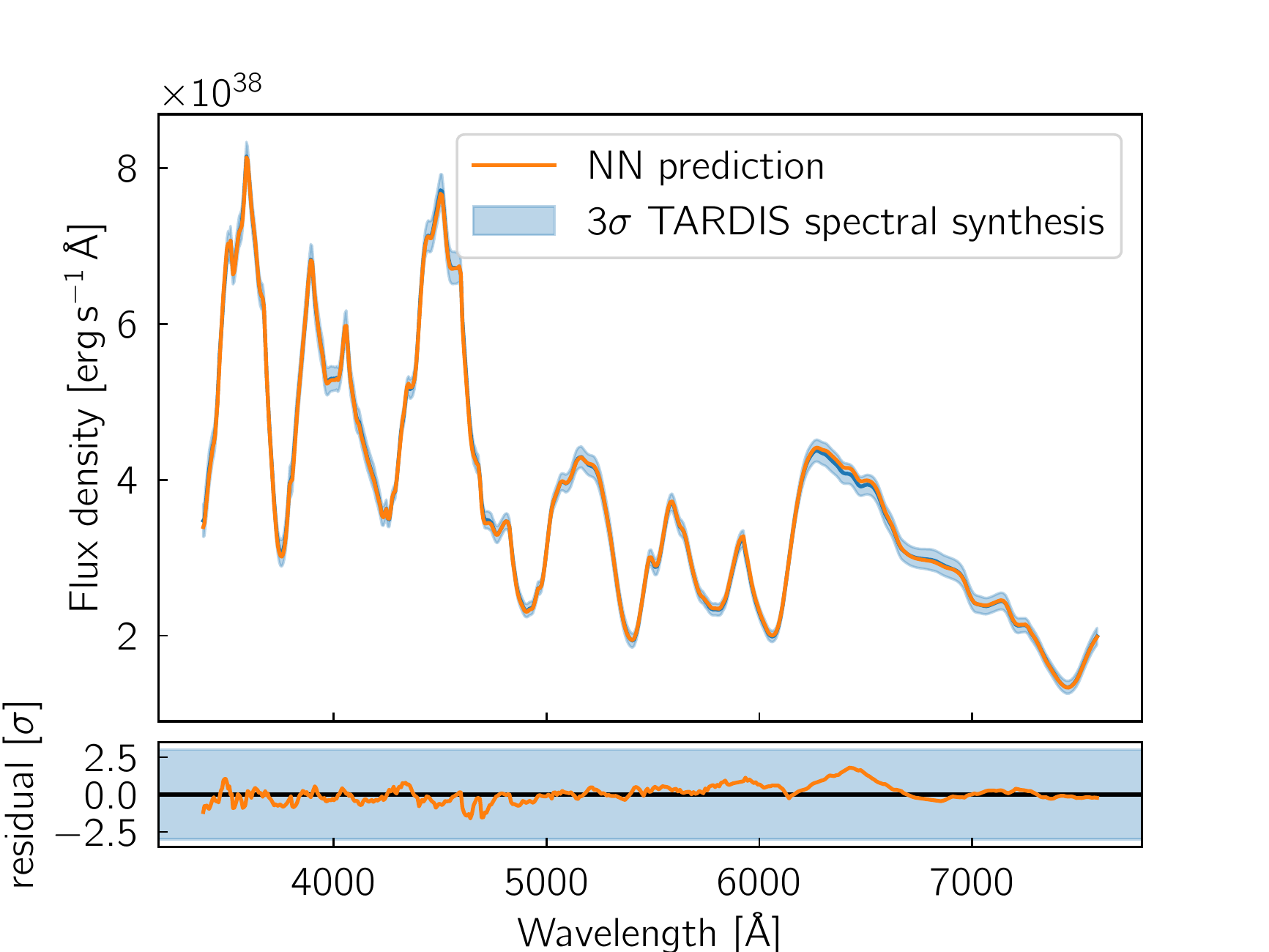}
    \caption{Comparison of the emulator prediction uncertainty with the Monte Carlo uncertainty.}
    \label{fig:tardis_emulator_uncertainty}
\end{figure}

We remind the reader that \tardis is based on an iterative Monte Carlo algorithm. The method results in variations in the final spectrum given different random seeds. We have run the worst predicting parameter set (see Figure~\ref{fig:best_worst_spectrum}) with 100 different seeds to test the variation. Figure~\ref{fig:tardis_emulator_uncertainty} shows that the prediction uncertainty of the emulator is close to the uncertainty of the Monte Carlo algorithm.  

For the desired application both MeanFE and MaxFE of the emulation will not contribute significantly as the systematic uncertainties will be much larger \citep[MeanFE for \sn{2002}{bo} 18\%; see Figure 5.5;][]{2011phdt.......324k}.

The main reason to use an emulator compared to \tardis itself is the speedup. The mean and standard deviation runtime for all \tardis runs during training set creation on a single CPU on the MSU HPCC cluster are $602\,\textrm{s}\pm186\,\textrm{s}$ with a minimum of 253\,s and a maximum of 2054\,s. Ensemble network evaluation takes $85\pm13.7$\,ms, which is several thousand times faster than the \tardis evaluation. A toy example of exploring likelihoods shows that a 20-dimensional problem needs 26 million evaluations \citep[see algorithm \textrm{radfriends} in Table~1 in ][]{2016S&C....26..383B}, which with the emulator is possible within $\approx 25$ days but not achievable without an emulator ($\approx420$\,years). This can be improved by forgoing ensemble modeling and taking a 10\% accuracy loss but having an evaluation time of $13.3\pm0.46$\,ms, which would do the exploration within $\approx4$\,days.


\section{Conclusion and Future Work}
\label{sec:conclusion}
We present a 12-dimensional emulator for the \tardis radiative transfer code. The emulator can predict the spectrum with an accuracy of on average 1\% with a speedup of almost \num{10000} in so-called ensemble mode and a speedup of almost \num{50000} with a marginally lower accuracy in single mode. A major part of the prediction uncertainty is likely not the emulator itself but noise from the Monte Carlo method of \tardis. The chosen parameter space is focused on the \sneia modeling. However, the general methodology can be applied to a much broader set of problems.

The presented emulator is useful for exploring single spectra with abundances that are uniform throughout the envelope. Initial fitting of supernova spectra including researching likelihoods that incorporate systematic uncertainties to account for the mismatch between \tardis and observed spectra is already underway. 

A complete reconstruction of an exploded object from spectral time series will have more than a hundred parameters. This will require the development of more complex emulators. For such parameter spaces, we will need to use more constraining priors when generating the training set. The authors have already experimented with various schemes to find a training set (e.g., drawing from Kernel Density estimates of the HESMA abundances) but such work is outside the current scope of exploring neural networks as function approximators for radiative transfer codes. 

We have shown that emulators enable the exploration of high-dimensional parameter spaces even with costly simulations. Such tools will be important assets for the data-rich era that astronomy is entering. 

\section*{Contributor Roles}

We use the CRT standard (see \url{https://casrai.org/credit/}) for reporting our contributor roles:

\begin{itemize}
    \item \textbf{Conceptualization} - Kerzendorf, Vogl
    \item \textbf{Data curation} - Kerzendorf
    \item \textbf{Formal Analysis} - Kerzendorf, PvdS
    \item \textbf{Investigation} - Kerzendorf, PvdS, Contardo, Buchner
    \item \textbf{Methodology} - Kerzendorf, PvdS
    \item \textbf{Resources} - PvdS, Kerzendorf
    \item \textbf{Software} - Kerzendorf, PvdS, Vogl, Williamson
    \item \textbf{Validation} - Kerzendorf, PvdS
    \item \textbf{Visualization} - Kerzendorf, PvdS
    \item \textbf{Writing---original draft} - Kerzendorf, PvdS
    \item \textbf{Writing---review \& editing} - Kerzendorf, PvdS, Vogl, Williamson, Contardo, Buchner
\end{itemize}

\section*{Acknowledgements}

This research made use of \tardis, a community-developed software
package for spectral synthesis in supernovae
\citep{2014MNRAS.440..387K, 2019A&A...621A..29V, kerzendorf_wolfgang_2020_3902923}.
The development of \tardis received support from the
Google Summer of Code initiative
and from ESA's Summer of Code in Space program. \tardis makes extensive use of Astropy and PyNE.

This work was supported in part through computational resources and services provided by the Institute for Cyber-Enabled Research at Michigan State University.
Neural network training was done at Volkswagen Group.
This work made use of the Heidelberg Supernova Model Archive (HESMA), \url{https://hesma.h-its.org}.

The authors would like to thank Gabriella Contardo for helpful with learning the neural network 

\bibliographystyle{aasjournal}
\bibliography{wekerzendorf}

\end{document}

%% file: glossary.tex
\PassOptionsToPackage{draft}{hyperref}
\usepackage{xspace}

\usepackage[xindy, toc, hyperfirst=false, nolist, nostyles, sanitize={name=false,description=false,symbol=false}]{glossaries}
\glsdisablehyper

\newglossaryentry{nlp}{name=NLP, description={Natural Language Processing}, first={Natural Language Processing (NLP)}}

\newglossaryentry{vrad}{name={radial velocity~}, text={radial velocity}, symbol={\ensuremath{v_\textrm{rad}}}, description={radial velocity}, sort=vrad}
\newglossaryentry{vrot}{name={stellar rotation~}, name={stellar rotation}, symbol={\ensuremath{v_\textrm{rot}}}, description={radial velocity}, sort=vrot}

\newcommand{\kms}{\ensuremath{\textrm{km}~\textrm{s}^{-1}}}

\newcommand{\xray}{X-ray}

\newcommand{\nucl}[2]{\ensuremath{^{#2}\textrm{#1}}}

\newcommand{\sn}[2]{SN~#1#2\xspace}

\newglossaryentry{angstrom}{name=\AA, description={unit of length $10^{-10}$\,m}, sort=angstrom}
\newglossaryentry{nir}{name=NIR,description={near infrared},first = {near infrared (NIR)}}
\newglossaryentry{psf}{name=PSF,description={point-spread function},first = {point-spread function (PSF)}}
\newglossaryentry{fwhm}{name=FWHM,description={Full Width Half Maximum},first = {FWHM}}
\newglossaryentry{rms}{name=RMS,description={Root Mean Square},first = {RMS}}
\newglossaryentry{signalnoise}{name=S/N,description={signal to noise}}
\newglossaryentry{uv}{name=UV,description={ultra violet},first = {ultra violet (UV)}}
\newglossaryentry{halpha}{name=\ensuremath{\textrm{H}\alpha}, description={First line of the Balmer series at 6563\,\AA}, sort=halpha}
\newglossaryentry{mgb}{name={Mg \textsc{i} b}, description={Triplet at 5167\,\AA, 5173\,\AA and 5184\,\AA}}
\newglossaryentry{sobolevapprox}{name={Sobolev approximation}, description={Lines are approximation with an infinitley thin interaction region \citep[e.g. no broadening][]{1960mes..book.....S}}, first={Sobolev approximation }}
\newglossaryentry{radeq}{name={radiative equilibrium}, description={The net flux of energy between matter and radiation field is zero}}
\newglossaryentry{nebularapprox}{name={nebular approximation}, description={Assumes that the plasma condition are controlled by a central radiation source. The radiation field decreases with the distance to the source by geometrical dilution. See \citet{1978stat.book.....M} for details}}
\newglossaryentry{modnebularapprox}{name={modified nebular approximation}, description={In contrast to \gls{nebularapprox} where only geometrical dilution is taken into account, the modified nebular approximation also takes dilution by other radiative processes into account }, first={modified nebular approximation}, parent=nebularapprox}
\newglossaryentry{thompsonscat}{name={Thomson scattering}, description={Scattering of photons on low energy electrons}}
\newglossaryentry{lte}{name={LTE}, description={Local Thermodynamic Equilibrium}, first={local thermodynamic equilibrium (LTE)}}
\newglossaryentry{lsr}{name={LSR}, description={Local Standard of Rest}, first={\textit{local standard of rest} (LSR)}}
\newglossaryentry{mc}{name={MC}, description={Monte Carlo}, first={\textit{Monte Carlo} (MC)}}
\newglossaryentry{wcs}{name={WCS}, description={world coordinate system}, first={world coordinate system (WCS)}}
\newglossaryentry{cmf}{name=CMF, text=CMF, first=Comoving Frame (CMF henceforth), description={Comoving Frame}}

\newglossaryentry{uvoir}{name=UVOIR, text=UVOIR, first=UV/optical/Near-IR (UVOIR), description={UV/optical/Near-IR}}
\newglossaryentry{ccd}{name=CCD,description={Charged Coupled Device}, first={charged coupled device (CCD)}, firstplural={charged coupled devices (CCDs)}}

\newglossaryentry{ew}{name=Equivalent Width, text={EW}, description={width of a rectangle that has the same area as a spectral line when taken to zero flux}, first={equivalent width (EW)}, firstplural={equivalent widths (EWs)}}
\newglossaryentry{agb}{name=AGB,description={Asymptotic Giant Branch}, first={Asymptotic Giant Branch (AGB)}}
\newglossaryentry{cmb}{name=CMB,description={Cosmic Microwave Background}}
\newglossaryentry{csm}{name=CSM,description={Circumstellar Medium}, first={circumstellar medium (CSM)}}
\newglossaryentry{csi}{name=CSI,description={Circumstellar Interaction}, first={circumstellar interaction (CSI)}}
\newglossaryentry{ism}{name=ISM,description={Interstellar Medium}, first={interstellar medium (ISM)}}
\newglossaryentry{ige}{name=IGE,description={Iron Group Element}, first={iron group element (IGE)}, firstplural={iron group elements (IGEs)}}
\newglossaryentry{epm}{name=EPM,description={Expanding Photosphere Method \citep{1974ApJ...193...27K}}, first={Expanding Photosphere Method (EPM)}}
\newglossaryentry{aic}{name=AIC,description={Accretion Induced Collapse}, first={accretion induced collapse (AIC)}}
\newglossaryentry{ime}{name=IME,description={Intermediate Mass Element}, first={intermediate mass element (IME)}, firstplural={intermediate mass elements (IMEs)}}
\newglossaryentry{h0}{name=\ensuremath{H_0},description={Hubbles constant}}
\newglossaryentry{nse}{name=NSE,description={Nuclear Statistical Equilibrium}, first={nuclear statistical equilibrium (NSE)}}
\newglossaryentry{cdm}{name=CDM,description={Cold Dark Matter}}
\newglossaryentry{grb}{name=GRB,description={Gamma Ray Burst}, first={Gamma Ray Burst (GRB)}, firstplural={Gamma Ray Bursts (GRBs)}}
\newglossaryentry{xps}{name=XPS, description={X-ray point source}, first={X-ray point source (XPS)}, firstplural={X-ray point sources (XPS)}}
\newglossaryentry{donor}{name=donor,description={non-degenerate companion in the \gls{sds}}}
\newglossaryentry{mainsequence}{name=main sequence,description={main sequence star}}
\newglossaryentry{redgiant}{name=red giant,description={red giant star}}
\newglossaryentry{mlcs}{name=MLCS,description={Multicolor Light Curve Shape method \citep[MLCS;][]{1996ApJ...473...88R}}, first={Multicolor Light-Curve Shape method \citep[MLCS;][]{1996ApJ...473...88R}}}
\newglossaryentry{rsoph}{name=RS~Ophiuci ,description={white dwarf accreting from a red giant - assumed progenitor of the \gls{sds}}, sort=rsoph}
\newglossaryentry{usco}{name=U~Scorpii,description={white dwarf accreting from a main sequence star - assumed progenitor of the \gls{sds}}, sort=usco}
\newglossaryentry{rcw86}{name=RCW~86,description={supernova remnant sometimes associated with \sn{185}{}}, sort=rcw86}
\newglossaryentry{casa}{name=Cas~A,description={Cassiopeia A supernova remnant - probably a \gls{snib} event}}
\newglossaryentry{cepheid}{name=Cepheid,description={very luminous variable star with a strong luminosity period relationship}}
\newglossaryentry{urca}{name=Urca, text=\textit{Urca}, description={process predominatly contributing to cooling in stars. The \textit{Urca} process consists of alternating electron-capture and $\beta^{-}$ decay of two nuclei pairs.},sort=urca}
\newglossaryentry{alphacen}{name=Alpha Centauri,description={one of the brightest stars in the night sky and a close binary}}
\newglossaryentry{pcygni}{name={P Cygni}, text={P Cygni},description={a hypergiant luminous blue variable with strong winds. Often referred to as a description for their line profiles showing a emission peak at the rest wavelength of the line and a blue-shifted absorption trough.}}

\newglossaryentry{teff}{name={effective temperature~}, text={effective temperature}, symbol={\ensuremath{T_\textrm{eff}}}, description={Temperature of a blackbody emitting the same total energy}, sort=teff}

\newglossaryentry{logg}{name={surface gravity~}, text={surface gravity}, symbol={\ensuremath{\textrm{log}\,g}}, description={gravity at the surface of a star}, sort=logg}
\newglossaryentry{feh}{name={metallicity~}, text={metallicity}, symbol=\textrm{[Fe/H]},description={iron abundance relative to the sun}, sort=feh}

\newglossaryentry{texp}{name={time since explosion~}, text={time since explosion}, text={time since explosion}, symbol={\ensuremath{t_{\rm exp}}},description={time since explosion (measured in days)}, sort=texp, first={time since explosion (\ensuremath{t_{\rm exp}})}}

\newglossaryentry{lmc}{name=LMC,description={Large Magellanic Cloud}, first={Large Magellanic Cloud (LMC)}, sort=lmc}
\newglossaryentry{smc}{name=SMC,description={Small Magellanic Cloud}, sort=smc}
\newglossaryentry{z}{name=\ensuremath{z},description={redshift}, sort=z}

\newglossaryentry{pca}{name=PCA,description={Princpal Component Analysis}, first={Princpal Component Analysis (PCA)}}

\newglossaryentry{gp}{name=GP,description={Gaussian Process}, first={Gaussian Process (GP)}}

\newglossaryentry{sfit}{name=SFIT, text=\textsc{sfit}, description={spectral fitting program for hot stars \citep{2001A&A...376..497J}}, first={\textsc{sfit} \citep{2001A&A...376..497J}}}
\newglossaryentry{iraf}{name=IRAF, text=\textsc{iraf}, description={Image Reduction and Analysis Facility maintained by NOAO}, first={\textsc{iraf}\protect\footnote{IRAF: the Image Reduction and Analysis Facility is distributed by the National Optical Astronomy Observatory, which is operated by the Association of Universities for Research in Astronomy (AURA) under cooperative agreement with the National Science Foundation (NSF).}}}
\newglossaryentry{pyraf}{name=PyRAF, text=\textsc{PyRAF}, description={Python wrap of \gls{iraf} maintained by STSCI}, first=\textsc{PyRAF} \protect\footnote{PyRAF is a product of the Space Telescope Science Institute, which is operated by AURA for NASA.}}
\newglossaryentry{astropy}{name=ASTROPY, text=\textsc{astropy}, description=\textsc{astropy} framework, first = \textsc{astropy} \citep{2013A&A...558A..33A}}
\newglossaryentry{numpy}{name=NUMPY, text=\textsc{numpy}, description=\textsc{numpy} framework, first = \textsc{numpy} \citep{walt2011numpy}}
\newglossaryentry{scipy}{name=SCIPY, text=\textsc{scipy}, description=\textsc{scipy} framework, first = \textsc{scipy} \citep{Jones:2001fk}}
\newglossaryentry{matplotlib}{name=matplotlib, text=\textsc{matplotlib}, description=\textsc{matplotlib} framework, first = \textsc{matplotlib} \citep{hunter2007matplotlib}}
\newglossaryentry{pandas}{name=pandas, text=\textsc{pandas}, description=\textsc{pandas} framework, first = \textsc{pandas} \citep{mckinney2010data}}
\newglossaryentry{ipython}{name=ipython, text=\textsc{ipython}, description=\textsc{ipython} framework, first = \textsc{ipython} \citep{perez2007ipython}}
\newglossaryentry{jupyter}{name=jupyter, text=\textsc{jupyter}, description=\textsc{jupyter} framework, first = \textsc{jupyter} \citep{kluyver2016jupyter,perez2015project,ragan2014jupyter}}
\newglossaryentry{aplpy}{name=aplpy, text=\textsc{aplpy}, description=\textsc{aplpy} framework, first = \textsc{aplpy} \citep{2012ascl.soft08017R}}
\newglossaryentry{nltk}{name=nltk, text=\textsc{nltk}, description=\textsc{nltk} framework, first = Natural Language ToolKit \citep[\textsc{NLTK};][]{bird2009natural}}
\newglossaryentry{scikit-learn}{name=scikit-learn, text=\textsc{scikit-learn}, description=\textsc{scikit-learn} framework, first = \textsc{scikit-learn} \citep[][]{scikit-learn}}
\newglossaryentry{scikit-image}{name=scikit-image, text=\textsc{scikit-image}, description=\textsc{scikit-image} framework, first = \textsc{scikit-image} \citep[][]{scikit-image}}
\newglossaryentry{moog}{name=MOOG,text={\textsc{moog}}, description={spectral synthesis software \citep{1973ApJ...184..839S}}, first={\textsc{Moog} \citep{1973ApJ...184..839S}}}
\newglossaryentry{atlas9}{name=ATLAS9,description={grid of stellar atmospheres \citep{2004astro.ph..5087C}}, first={ATLAS9 \citep{2004astro.ph..5087C}}}
\newglossaryentry{vald}{name=VALD,description={Vienna Atomic Line Database \citep{2000BaltA...9..590K}}, first={Vienna Atomic Line Database \citep[VALD;][]{2000BaltA...9..590K}}}
\newglossaryentry{sextractor}{name=SExtractor, text=\textsc{SExtractor}, description={Source Extractor photometry program \citep{1996A&AS..117..393B}}, first={\textsc{SExtractor} \citep{1996A&AS..117..393B}}}
\newglossaryentry{swarp}{name=SWarp, text=\textsc{SWarp}, description={SWarp \citep{2002ASPC..281..228B}}, first={\textsc{SWarp} \citep{2002ASPC..281..228B}}}
\newglossaryentry{astrometry.net}{name=astrometry.net, text=\textsc{astrometry.net}, description={\textsc{astrometry.net} \citep{2010AJ....139.1782L}} first={\textsc{astrometry.net} \citep{2010AJ....139.1782L}}}

\newglossaryentry{astrodrizzle}{name=AstroDrizzle, text=\textsc{AstroDrizzle}, description={AstroDrizzle \citep{2012drzp.book.....G}}, first={\textsc{AstroDrizzle} \citep{2012drzp.book.....G}}}

\newglossaryentry{idl}{name=IDL,text={\textsc{idl}}, description={Interactive Data Language}}
\newglossaryentry{makee}{name=MAKEE,text=\textsc{makee}, description={MAuna Kea Echelle Extraction by Tom Barlow available}}
\newglossaryentry{minuit}{name=MINUIT,text={\textsc{minuit}}, description={collection of numerical optimization tools \citep{James:1975dr}}}
\newglossaryentry{migrad}{name=MIGRAD,text={\textsc{migrad}}, description={numerical gradient optimization tools - part of \gls{minuit}}}
\newglossaryentry{dolphot}{name=DOLPHOT, text=\textsc{dolphot}, description=photometry package for HST, first=\textsc{dolphot} \citep{2000PASP..112.1383D}}
\newglossaryentry{synphot}{name=synphot, text={\textsc{synphot}}, description={synthetic photometry package from STSCI}, first={\textsc{synphot}\protect\footnote{\textsc{synphot} is a product of the Space Telescope Science Institute, which is operated by AURA for NASA.}}}
\newglossaryentry{chianti}{name=CHIANTI, text=CHIANTI, description= CHIANTI Database 7.1, first =CHIANTI 7.1 \citep{1997A&AS..125..149D,2012ApJ...744...99L}}
\newglossaryentry{synpp}{name=SYNPP, text=SYN++, description= SYN++ software, first =SYN++ \citep{2011PASP..123..237T}}
\newglossaryentry{tardis}{name=TARDIS, text=\textsc{tardis}, description= TARDIS MC code, first = {\textsc{tardis} \citep{2014MNRAS.440..387K}}}

\newglossaryentry{artis}{name=ARTIS, text=\textsc{artis}, description= ARTIS MC code, first = \textsc{artis} \citep{2009MNRAS.398.1809K}}
\newglossaryentry{cmfgen}{name=CMFGEN, text=\textsc{cmfgen}, description=CMFGGEn radiative transfer code, first = \textsc{cmfgen} \citep{1998ApJ...496..407H}}
\newglossaryentry{sedona}{name=SEDONA, text=\textsc{sedona}, description= Sedona MC code, first = \textsc{sedona} \citep{2006ApJ...651..366K}}
\newglossaryentry{phoenix}{name=PHOENIX, text=\textsc{phoenix}, description= PHOENIX radiative transfer code, first = \textsc{phoenix} \citep{1999jcoam.109...41h,1998ApJ...495..370B,1997ApJ...483..390H,1996ApJ...462..386H,1997ApJ...490..803H}}
\newglossaryentry{mlmc}{name=MLMC, text=ML93, description= Mazzali Lucy Monte Carlo, first ={Mazzali \& Lucy (1993, ML93) code}}
\newglossaryentry{starkit}{name=STARKIT, text=\textsc{starkit}, description= TARDIS MC code, first = {\textsc{starkit} \citep{wolfgang_kerzendorf_2015_28016}}}

\newglossaryentry{pyne}{name=PYNE, text=\textsc{pyne}, description= PYNE code, first = {\textsc{pyne} \citep{Scopatz2012a}}}
\newglossaryentry{multinest}{name=MULTINEST, text=\textsc{MultiNest}, description=MultiNest, first={\textsc{MultiNest} \citep{2009MNRAS.398.1601F}}}
\newglossaryentry{wsynphot}{name=WSYNPHOT, text=\textsc{wsynphot}, description=Wsynphot, first={\textsc{wsynphot}\protect\footnote{\protect\url{https://github.com/wkerzendorf/wsynphot}}}}
\newglossaryentry{specutils}{name=SPECUTILS, text=\textsc{specutils}, description=specutils, first={\textsc{specutils} \protect\footnote{\protect\url{https://github.com/astropy/specutils}}}}
\newglossaryentry{ads}{name=ADS ,description=ADS, first={NASA Astrophysics Data System (ADS) \citep{2000A&AS..143...41K}}}

\newglossaryentry{2mass}{name=2MASS,description={Two Micron All Sky Survey \citep{2006AJ....131.1163S}}, first={Two Micron All Sky Survey \citep{2006AJ....131.1163S}}}
\newglossaryentry{wiserep}{name=\textsc{WISeREP}, description={Weizmann Interactive Supernova data REPository \citep{2006AJ....131.1163S}}, first={\textsc{WISeREP} \citep{2012PASP..124..668Y}}}
\newglossaryentry{nomad}{name=NOMAD,first={Naval Observatory Merged Astrometric Dataset \citep[NOMAD; ][]{2005yCat.1297....0Z}}, description={Naval Observatory Merged Astrometric Dataset}}
\newglossaryentry{sdss}{name=SDSS, description={Sloan Digital Sky Survey}}
\newglossaryentry{dss}{name=DSS, description={Digitized Sky Survey}}

\newglossaryentry{eso}{name=ESO, description={European Southern Observatory}, first={European Southern Observatory (ESO)}}
\newglossaryentry{eso.opc}{name=OPC, description={Observing Programmes Comittee}, first={Observing Programmes Comittee (OPC)}}
\newglossaryentry{iau}{name=IAU,description={International Astronomical Union}, first={IAU}}
\newglossaryentry{ctio}{name= CTIO, description={Cerro Tololo Inter-American Observatory}, first={Cerro Tololo Inter-American Observatory (CTIO)}}

\newglossaryentry{nsf}{name=NSF, description={National Science Foundation}, first={National Science Foundation (NSF)}}

\newglossaryentry{wifes}{name=WIFES, text=\textsc{WiFeS}, first={\textsc{WiFeS} \citep{2007Ap&SS.310..255D}},  description={Wide Field Spectrograph - \gls{ifu} mounted on the 2.3\,m telescope at Siding Spring Observatory}}

\newglossaryentry{scp}{name=SCP, description={Supernova Cosmology Project, led by Saul Perlmutter}, first={Supernova Cosmology Project (SCP)}}
\newglossaryentry{hzsns}{name=HZSNS, description={High Z Supernova Search, led by Brian Schmidt}, first={High Z Supernova Search (HZSNS)}}

\newglossaryentry{vlt}{name=VLT,description={Very Large Telescope located on Cerro Paranal (Chile)}, first={Very Large Telescope (VLT)}}
\newglossaryentry{flames}{name=FLAMES,description={Multi-object, intermediate and high resolution spectrograph mounted on the  \gls{vlt}}}

\newglossaryentry{hires}{name=HIRES, description={High Resolution Echelle Spectrometer mounted on the Keck Telescope}, first={High Resolution Echelle Spectrometer \citep[HIRES;][]{1994SPIE.2198..362V}}}

\newglossaryentry{lris}{name=LRIS,description={Low Resolution Imaging Spectrometer mounted on the Keck Telescope}, first={Low-Resolution Imaging Spectrometer \citep[LRIS;][]{Oke95}}}

\newglossaryentry{decam}{name=DECam, description={DECam is a high-performance, wide-field CCD imager mounted at the prime focus of the Blanco 4-m telescope at \gls{ctio}.}, first={Dark Energy Camera \citep[DECam; ][]{2012PhPro..37.1332D,2015AJ....150..150F}}}

\newglossaryentry{essence}{name=ESSENCE,description={The `Equation of State: SupErNovae trace Cosmic Expansion' project \citep[ESSENCE;][]{2002AAS...201.7809G}}, first={`The Equation of State: SupErNovae trace Cosmic Expansion' \citep[ESSENCE;][]{2002AAS...201.7809G}}}
\newglossaryentry{ifu}{name=IFU,description={Optical instrument combining spectrographic and imaging capabilities, used to obtain spatially resolved spectra}, first={Integral Field Unit (IFU)}, firstplural={Integral Field Units (IFUs)}}

\newglossaryentry{besancon}{name=Besan\c{c}on Model, description={Model of stellar population synthesis of the Galaxy, including kinematics.}}

\newglossaryentry{int}{name=INT,description={Isaac Newton 2.5\,m Telescope}, first={Isaac Newton 2.5\,m Telescope (INT)}}

\newglossaryentry{chandra}{name=Chandra,description={Chandra \xray\ Observatory (space-based)}}
\newglossaryentry{hst}{name=HST,description={Hubble Space Telescope}}
\newglossaryentry{hst.wfpc2}{name=WFPC2,description={Wide-Field Planetary Camera 2 mounted on the \gls{hst}}, first={Wide-Field Planetary Camera 2 (WFPC2)}}
\newglossaryentry{hst.acs}{name=ACS,description={Advanced Camera for Surveys mounted on the \gls{hst}}, first={Advanced Camera for Surveys (ACS)}}
\newglossaryentry{hst.wfc3}{name=WFC3,description={Wide-Field Camera 3 mounted on the \gls{hst}}, first={Wide-Field Camera 3 (WFC3)}}
\newglossaryentry{hst.cte}{name=CTE, description={charge transfer efficiency (CTE)}, first={charge transfer efficiency \citep[CTE; see ][for a description]{2009acs..rept....1C}}}

\newglossaryentry{snls}{name=SNLS,description={Supernova Legacy Survey \citep{2003AAS...203.8209P}}, first={Supernova Legacy Survey \citep[SNLS;][]{2003AAS...203.8209P}}}
\newglossaryentry{dass}{name=DASS, description={Digitized Astronomy Supernova Survey \citep{1975PASP...87..565C}}, first={Digitized Astronomy Supernova Survey \citep[DASS;][]{1975PASP...87..565C}}}
\newglossaryentry{bait}{name=BAIT, description={Berkley Automatic Imaging Telescope \citep{1993PASP..105.1164R}}, first={Berkley Automatic Imaging Telescope \citep[BAIT;][]{1993PASP..105.1164R}}}
\newglossaryentry{kait}{name=KAIT, description={Katzman Automatic Imaging Telescope \citep{2001ASPC..246..121F}}, first={Katzman Automatic Imaging Telescope \citep[KAIT;][]{2001ASPC..246..121F}}}
\newglossaryentry{loss}{name=LOSS, description={Lick Observatory Supernova Search  \citep{2000AIPC..522..103L}}, first={Lick Observatory Supernova Search \citep[LOSS;][]{2000AIPC..522..103L}}}
\newglossaryentry{ctss}{name=CTSS,description={Cal\'{a}n/Tololo Supernova Survey \citep{1993AJ....106.2392H}}, first={Cal\'{a}n/Tololo supernova survey \citep[CTSS;][]{1993AJ....106.2392H}}}

\newglossaryentry{ptf}{name=PTF, description={Palomar Transient Factory \citep{2009PASP..121.1334R}}, first={Palomar Transient Factory \citep[PTF;][]{2009PASP..121.1334R}}}
\newglossaryentry{batse}{name=BATSE, description={Burst and Transient Source Experiment mounted on the Compton Gamma Ray Observatory}, first={Burst and Transient Source Experiment (BATSE)}}
\newglossaryentry{bepposax}{name=BeppoSAX, description={\xray\ satellite named in honor of Giuseppe "Beppo" Occhialini}}
\newglossaryentry{rosat}{name=ROSAT, description={short for R\"{o}ntgensatellit}, first={ROSAT}}
\newglossaryentry{hete2}{name=HETE2, description={High Energy Transient Explorer}, first={High Energy Transient Explorer (HETE)}}
\newglossaryentry{ska}{name=SKA, description={Square Kilometre Array}, first={Square Kilometre Array (SKA)}}
\newglossaryentry{swift}{name=Swift, description={Swift Gamma-Ray Burst Mission}}

\newglossaryentry{gnirs}{name=GNIRS, description={Gemini Near InfraRed Spectrograph mounted on the Gemini North Telescope}}
\newglossaryentry{gmosn}{name=GMOS, description={Gemini Multi Object Spectrograph mounted on the
 Gemini North Telescope}, first={GMOS \citep[Gemini Multi Object Spectrograph;][]{2004PASP..116..425H}}}

\newglossaryentry{vla}{name=VLA, description={Very Large Array radio telescope located in North America}, first={Very Large Array (VLA)}}
\newglossaryentry{evla}{name=EVLA, description={Extended Very Large Array radio telescope located in North America}, first={Extended Very Large Array (EVLA)}}
\newglossaryentry{skymapper}{name=SkyMapper, description={SkyMapper telescope \citep{2007PASA...24....1K}}, first={SkyMapper \citep{2007PASA...24....1K}}}
\newglossaryentry{panstarrs}{name=PanSTARRS, description={Panoramic Survey Telescope \& Rapid Response System \citep{2004SPIE.5489...11K}}, first={Panoramic Survey Telescope \& Rapid Response System \citep[PanSTARRS;][]{2004SPIE.5489...11K}}}
\newglossaryentry{ps1dr1}{name=PS1~DR1, description={Panoramic Survey Telescope \& Rapid Response System \citep{2004SPIE.5489...11K} }, first={Panoramic Survey Telescope \& Rapid Response System \citep[PanSTARRS;][]{2004SPIE.5489...11K} DR1}}

\newglossaryentry{lsst}{name=LSST, description={Large Synoptic Survey Telescope}, first={Large Synoptic Survey Telescope \citep[LSST;][]{2006AAS...209.8604P}}}
\newglossaryentry{ppmxl}{name=PPMXL, description={PPMXL Catalog of Positions and Proper Motions on the ICRS \citep{2010AJ....139.2440R}}}
\newglossaryentry{gaia}{name=GAIA, description={Global Astrometric Interferometer for Astrophysics \citep{2001A&A...369..339P}}, first={Global Astrometric Interferometer for Astrophysics \citep[GAIA;][]{2001A&A...369..339P}}}
\newglossaryentry{ligo}{name=LIGO, description={Laser Interferometer Gravitational Wave Observatory}, first={Laser Interferometer Gravitational Wave Observatory \citep[LIGO;][]{1992Sci...256..325A}}}
\newglossaryentry{aligo}{name=Advanced LIGO, description={Advanced LIGO}, sort=ligo2}
\newglossaryentry{lisa}{name=LISA, description={Laser Interferometer Space Antenna \citep{1994ESAJ...18..219J}}, first={Laser Interferometer Space Antenna \citep[LISA;][]{1994ESAJ...18..219J}}}

\newglossaryentry{irc}{name=IRC, text={IRC}, description={infrared catastrophe}, first={infrared catastrophe \citep[IRC;][]{1980PhDT.........1A}}}

\newglossaryentry{sn}{name=Supernova, text={SN}, plural={SNe}, description={exploding star}, nonumberlist=true, first={supernova (SN)}, firstplural={supernovae (SNe)}}
\newglossaryentry{snia}{name=Type~Ia (SN~Ia), text={SN~Ia}, description={Thermonuclear explosion of a white dwarf - spectra show no hydrogen but a strong silicon line},first={Type~Ia supernova (SN~Ia)}, firstplural={Type Ia supernovae (SNe~Ia)}, plural={SNe~Ia}, parent=sn, nonumberlist=true}
\newcommand{\sneia}{\glspl*{snia}\xspace}
\newcommand{\snia}{\gls*{snia}\xspace}

\newglossaryentry{branchnormal}{name={branch-normal}, text=\textit{Branch-normal}, description={Large homogeneous class of Type Ia Supernovae, defined in \citet{1993AJ....106.2383B}}, first={\textit{Branch-normal} SNe Ia \citep{1993AJ....106.2383B}}, parent=snia}
\newglossaryentry{91t}{name={91T-like}, description={Luminous class of Type Ia supernovae similar to \sn{1991}{T} \citep{1992AJ....103.1632P}} , first={91T-like}, parent=snia}
\newglossaryentry{91bg}{name={91bg-like}, description={Faint class of Type Ia supernovae similar to \sn{1991}{bg} \citep{1992AJ....104.1543F}}, first={91bg-like}, parent=snia}
\newglossaryentry{02cx}{name={02cx-like}, description={Peculiar class of Type Ia supernovae similar to \sn{2002}{cx} \citep{2003PASP..115..453L}}, first={02cx-like \sneia\ \citep{2003PASP..115..453L}}, parent=snia}

\newglossaryentry{snibc}{name=Type~Ib/c, text={SN~Ib/c}, description={Collapse of the core of a massive star -  spectrum shows no hydrogen and no silicon line},first={Type~Ib/c supernova (SN~Ib/c)}, firstplural={Type~Ib/c supernovae (SNe~Ib/c)}, plural={SNe~Ib/c}, parent=sn}

\newglossaryentry{snib}{name=Type~Ib, text={SN~Ib}, description={Spectrum shows no hydrogen and no silicon, but helium line},first={Type Ib supernova (SN~Ib)}, firstplural={Type~Ib supernovae (SNe~Ib)}, plural={SNe~Ib}, parent=snibc}

\newglossaryentry{snic}{name=Type~Ic, text={SN~Ic}, description={Spectrum shows no hydrogen, no silicon and no helium line},first={Type~Ic supernova (SN~Ic)}, firstplural={Type~Ic supernovae (SNe~Ic)}, plural={SNe~Ic}, parent=snibc}

\newglossaryentry{snii}{name=Type~II, text={SN~II}, description={Collapse of the core of a massive star - spectrum shows strong hydrogen line},first={Type~II supernova (SN~II)}, firstplural={Type~II supernovae (SNe~II)}, plural={SNe~II}, parent=sn}

\newglossaryentry{sniib}{name=Type~IIb, text={SN~IIb}, description={Spectrum shows hydrogen and helium lines},first={Type~IIb supernova (SN~IIb)}, firstplural={Type~IIb supernovae (SNe~IIb)}, plural={SNe~IIb}, parent=snii}

\newglossaryentry{sniip}{name=Type~II~Plateau (Type IIP), text={SN~IIP}, description={Lightcurve shows plateau},first={Type~IIP supernova (SN~IIP)}, firstplural={Type~II Plateau supernovae \citep[SNe~IIP;][]{1979A&A....72..287B}}, plural={SNe~IIP}, parent=snii}

\newglossaryentry{sniil}{name=SN~II~Linear, text={SN~IIL}, description={Lightcurve shows no plateau, but linear decline},first={Type~IIL supernova (SN~IIL)}, firstplural={Type~II~Linear supernovae \citep[SNe~IIL;][]{1990MNRAS.244..269S}}, plural={SNe~IIL}, parent=snii}

\newglossaryentry{sniin}{name=Type II narrow-lined (Type IIn), description={Spectrum shows narrow lines},first={Type~II~narrow-lined supernova (SN IIn)}, firstplural={Type~IIn supernovae (SNe~IIn)}, plural={SNe~IIn}, parent=snii}

\newglossaryentry{snr}{name=Remnant (SNR), text=SNR, description={Remnant left visible post-explosion}, first={supernova remnant (SNR)}, firstplural={supernova remnants (SNRs)}, parent=sn}

\newglossaryentry{dtd}{name=DTD,description={delay time distribution - expected supernova rate over time after a brief outburst of starformation},first={delay time distribution (DTD)}, firstplural={delay time distributions (DTDs)}, plural=DTDs}

\newglossaryentry{hvg}{name=HVG,description={high velocity gradient - Type Ia supernovae with a fast evolution of photospheric velocity},first={high velocity group (HVG)}, firstplural={high velocity groups (HVGs)}, plural=HVGs, parent=snia}

\newglossaryentry{lvg}{name=LVG,description={low velocity gradient - Type Ia supernovae with a slow evolution of photospheric velocity},first={low velocity group (LVG)}, firstplural={low velocity groups (LVGs)}, plural=LVGs, parent=snia}

\newglossaryentry{wd}{name=white dwarf (WD), text=WD, description={White Dwarf - extremely dense stellar remnant}, first={white dwarf (WD)}}
\newglossaryentry{onemgwd}{name= Oxygen/Neon (ONe), text={ONe-WD},description={Oxygen/Neon White Dwarf}, first={oxygen/neon White Dwarf (ONe-WD)}, parent=wd}
\newglossaryentry{cowd}{name=carbon/oxygen (CO), text={CO-WD}, description={carbon/oxygen white dwarf}, first={carbon/oxygen white dwarf (CO-WD)}, firstplural = {carbon/oxygen white dwarfs (CO-WDs)}, parent=wd}

\newglossaryentry{sds}{name=SD-Scenario,description={single-degenerate scenario (single white dwarf accreting from non-degenerate companion)}, first={single-degenerate scenario (SD-scenario)}}

\newglossaryentry{dds}{name=DD-Scenario, description={double degenerate scenario (merging of two white dwarfs)}, first={double-degenerate scenario (DD-scenario)}}

\newglossaryentry{sss}{name=SSS, text={supersoft \xray\ source}, description={supersoft \xray\ source - believed to be emitted by nuclear fusion on a white dwarf's surface}}

\newglossaryentry{amcvn}{name=AM CVn, description={AM Canum Venaticorum star \citep[white dwarf accreting hydrogen poor matter from a companion star; see ][]{2005ASPC..330...27N}}}

\newglossaryentry{rlof}{name=RLOF, description={Roche Lobe Overflow (see \citet{1971ARA&A...9..183P} for a more detailed description)}, first={Roche-lobe overflow (RLOF)}}

\newglossaryentry{mchan}{name={Chandrasekhar mass~}, text={Chandrasekhar~mass}, symbol={\ensuremath{M_\textrm{Chan}}}, plural={Chandrasekhar~masses}, description={Mass when the core of a star collapses due to insufficient degeneracy pressure - for a white dwarf $\approx1.38\,M_\odot$ see \citet{1931ApJ....74...81C}}, first={Chandrasekhar~mass \citep[$M_\textrm{Chan}=1.38\,M_\odot$;][]{1931ApJ....74...81C}}, sort=mchan}

\newglossaryentry{w7}{name={W7 model},description={W7 model \citep{1984ApJ...286..644N}},first = {W7 model \citep{1984ApJ...286..644N}}}

\newglossaryentry{stats.pdf}{name=PDF, description={Probability Density Function}, first={Probability Density Function}}

\newglossaryentry{dpr}{name=DPR, description={Distributed Peer Review}, first={Distributed Peer Review (DPR)}}

%% file: training_grid_description.tex
\begin{tabular}{lrrrrrrrrrr}
\toprule
{} &       C &    O &      Mg &    Si &     S &     Ca &      Ti &      Cr &    Fe &  $^{56}$Ni \\
\midrule
min  & 6.6e-06 & 0.05 & 2.5e-05 & 0.031 & 0.012 & 0.0016 & 3.6e-06 & 0.00019 & 0.005 &      0.025 \\
25\% & 8.6e-05 & 0.51 & 0.00014 & 0.056 & 0.022 &  0.003 & 6.7e-06 & 0.00031 & 0.011 &      0.052 \\
50\% &  0.0011 & 0.63 & 0.00071 &   0.1 &  0.04 & 0.0055 & 1.2e-05 & 0.00049 & 0.023 &       0.11 \\
75\% &   0.013 & 0.73 &  0.0038 &  0.19 & 0.071 &   0.01 & 2.2e-05 & 0.00078 &  0.05 &       0.22 \\
max  &    0.16 & 0.92 &   0.021 &  0.34 &  0.13 &  0.018 & 4.1e-05 &  0.0012 &  0.11 &       0.46 \\
\bottomrule
\end{tabular}

%% file: hyper_param_table.tex
\medskip\noindent\begin{tabular}{ll}
    \textbf{parameter}  & \textbf{values 
    }\\\hline
    \# hidden layers    & 2--6 \\
    \# neurons/layer    & 100--500 in steps of 100\\
    batch size          & 100, 500, 1000, 2000\\
    activation function & tanh, relu, selu, elu, softplus \\
    optimiser           & adam, nadam, adadelta, adagrad\\
    dropout rate        & 0--0.6 in steps of 0.2\\
    batch normalisation & after each layer / not at all\\
    initialiser         & glorot\_normal, he\_normal
    \end{tabular}\medskip
    
    \caption{\label{tab:hyper_param} A short explanation of the hyperparameters.
    Activation functions: tanh is a known mathematical function which keeps the output of the neuron between $+1$ and $-1$; relu (`rectified linear unit') equals $\mathrm{relu}(x)\equiv\max(0,x)$. softplus is a smooth version of relu and defined as $\ln(1+\exp(x))$.  The selu function is a normalized relu \citep{DBLP:journals/corr/KlambauerUMH17}.  The elu \citep{clevert2015fast} goes negative with $a(\exp(x)-1)$ for $x<0$.
    \\
    adadelta \citep{zeiler2012adadelta}, adagrad \citep{JMLR:v12:duchi11a}, adam \citep{kingma2014adam} are modern second-order optimisation methods used in neural network training.  nadam is adam but with Nesterov gradients \citep{sutskever2013importance}.
    \\
    Batch normalization \citep{ioffe2015batch} normalises the activations of a layer of neurons per batch and helps much in preventing overfitting.
    \\
    Dropout \citep{hinton2012improving} prevents overfitting by randomly switching hidden units off during training by the given rate. We never combined batch normalization with dropout.
    \\
    Early stopping is always done, by selecting that step in the optimization that has a low error on the cross-validation set.
    }

%% file: ensemble_list.tex
\begin{tabular}{llll}
\toprule
depth & optimizer & activation & width \\
\midrule
    4 &     nadam &   softplus &   200 \\
    4 &      adam &   softplus &   200 \\
    3 &      adam &   softplus &   300 \\
    3 &     nadam &   softplus &   400 \\
    4 &     nadam &   softplus &   200 \\
\bottomrule
\end{tabular}